\def\Rwd     {R_{\rm wd}}
\def\Mwd     {M_{\rm wd}}
\def\Msun    {{\rm M_{\odot}}}
\def\Mdot    {\mathaccent 95 M}
\newcommand\lax{\>\vcenter{\hbox{$<$\hskip-.75em\lower1.0ex\hbox{$\sim$}}}\>}
\newcommand\gax{\>\vcenter{\hbox{$>$\hskip-.75em\lower1.0ex\hbox{$\sim$}}}\>}

 \documentstyle[11pt,paspconf,epsf]{article}	

\begin{document}

\title{The EUV and X-ray Emission of Nonmagnetic Cataclysmic Variables}
\author{Christopher W.\ Mauche}
 \affil{Lawrence Livermore National Laboratory,\\
        L-41, P.O.~Box 808, Livermore, CA 94550}

\vbox to -10pt{\vskip -6.0cm
\hbox to \hsize{{\it 1997,  Proceedings of the 13th North American Workshop
                on Cataclysmic}}
\hbox to \hsize{{\it Variables, ed.\ S.~Howell, E.~Kuulkers, \& C.~Woodward
                (San Francisco: ASP)}}
\vss}
\vskip -10pt

\begin{abstract}
Recent results are presented and discussed regarding the EUV and X-ray emission
of nonmagnetic cataclysmic variables. Emphasis is given to high accretion rate
systems (novalike variables and dwarf novae in outburst), and to a number of
apparent discrepancies between observations and the theory of the boundary
layer between the accretion disk and the surface of the white dwarf. Discussed
are EUV and X-ray light curves, dwarf nova oscillations, and spectra, with
new and previously unpublished results on SS~Cyg and OY~Car.
\end{abstract}


\section{Introduction}

In 1985 Patterson \& Raymond presented a simple and appealing picture of the
X-ray emission of nonmagnetic CVs which for many has become the ``standard
model'' (Patterson \& Raymond 1985a; b). The essence of this model is embodied
in Figure 8 of Patterson \& Raymond (1985a): (1) the X-ray emission is produced
in the boundary layer between the accretion disk and the surface of the white
dwarf; (2) when the accretion rate $\Mdot $ onto the white dwarf is low (e.g.,
dwarf novae in quiescence), the boundary layer is optically thin and quite hot;
(3) when $\Mdot $ onto the white dwarf is high (e.g., novalike variables and
dwarf novae in outburst), the boundary layer is optically thick and quite cool;
and (4) even when $\Mdot $ is high, there invariably will be a hot, optically
thin ``atmosphere'' on the otherwise cool boundary layer. At one extreme, the
boundary layer can be as hot as the virial temperature $kT_{\rm vir} = G\Mwd
\mu m_{\rm H}/3\Rwd \sim 50$ keV, and, at the other extreme, as cool as the
blackbody temperature $kT_{\rm bl} = k(G\Mwd\Mdot/8\pi\sigma\Rwd ^3)^{1/4}\sim
10$~eV. The boundary layer luminosity is $L_{\rm bl}\approx 0.5\times \zeta
\times  G\Mwd\Mdot/\Rwd = 8\times 10^{34}\, \zeta\, (\Mwd/\Msun)
(\Mdot/10^{-8}\, \Msun\, {\rm yr}^{-1}) (\Rwd/5\times 10^8~{\rm cm})^{-1}~\rm
erg~s^{-1}$, where the factor $\zeta $ accounts for the rotation of the white
dwarf: $\zeta\equiv [1-(v/v_{\rm break})]^2$, where $v$ is the rotation
velocity and $v_{\rm break} = (G\Mwd/\Rwd)^{1/2}$ is the breakup velocity of
the white dwarf. The boundary layer luminosity is reduced significantly only
if $v/v_{\rm break}\gax 0.3$ or $v\gax 1600\, (\Mwd/\Msun)^{1/2} (\Rwd/5\times
10^8~{\rm cm})^{-1/2}~\rm km~s^{-1}$, whereas $v\,\sin i < 200$ km $\rm s^{-1}$
for U~Gem (Sion et~al.\ 1994), $\approx 600~\rm km~s^{-1}$ for VW~Hyi (Sion
et~al.\ 1995), and possibly $\approx 300~\rm km~s^{-1}$ for SS~Cyg (Mauche
1997c).

Narayan \& Popham (1993) and Popham \& Narayan (1995) have recently supplied
more detailed theoretical treatments of the boundary layers of CVs. In
high-$\Mdot $ systems, Popham \& Narayan distinguish between the ``dynamical
boundary layer'' very near the surface of the white dwarf where the disk
material switches from rotation to pressure support and the angular
velocity deviates significantly from Keplerian, and the much more extended
``thermal boundary layer'' where the boundary layer luminosity is radiated.
The structure of the boundary layer and the resulting radiation spectrum
depend on the mass-accretion rate and the white dwarf's mass and rotation
velocity; for $\Mwd =0.6$--$1.0\,\Msun $ the peak effective temperatures are
$kT_{\rm eff}\approx 17$--30~eV, but this can be reduced further if $v/v_{\rm 
break}\gax 0.1$.

To test these and other theoretical models of the boundary layers of CVs, we
present and discuss below some of the observational aspects of the EUV and
X-ray emission of nonmagnetic CVs. For another recent review, see Verbunt
(1996).

\section{Light Curves}

The physical location of the source of the X-ray emission in nonmagnetic CVs is
constrained by (1) the variation of apparent emission measure with inclination
(van Teeseling et~al.\ 1996), (2) the relative strength of the line and
continuum spectrum reflected off the white dwarf and accretion disk (e.g., 
Done \& Osborne 1997), and (3) the X-ray light curves of eclipsing systems. All
these constraints imply that all or most of the hard X-rays are emitted very
close to the white dwarf, but the eclipse observations are the most direct and
unambiguous. In Z~Cha ($i\approx 82^\circ $; van Teeseling 1997a) and HT~Cas
($i\approx  81^\circ $; Wood et~al.\ 1995a; Mukai et~al.\ 1997) in quiescence,
the X-rays are fully eclipsed for $\Delta\phi\sim 0.04$, comparable to the
length of the optical eclipse of the white dwarf. From the high-quality
{\it ASCA\/} X-ray light curve of HT~Cas (Fig.~1), the duration of the
ingress/egress is measured to be $\Delta\phi\approx 0.004$, implying that the
size of the X-ray emission region $R_{\rm X}\lax 1.15\,\Rwd $.

\begin{figure}
\epsfbox[130 530 475 700]{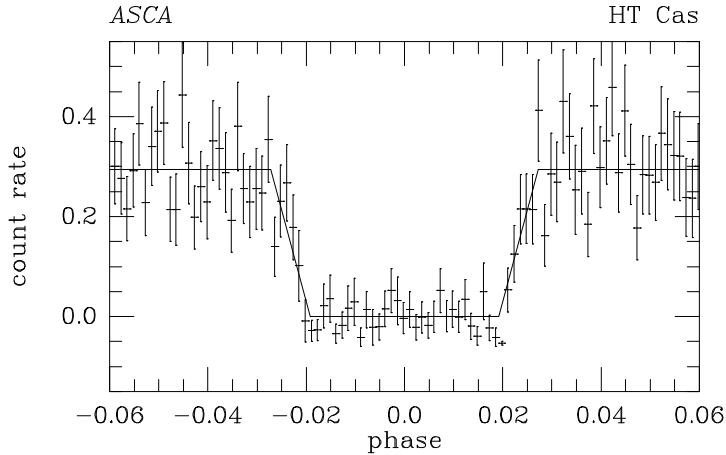}	
\caption{Folded {\it ASCA\/} light curve of HT~Cas in quiescence (after Mukai
et~al.\ 1997).}
\label{fig1}
\end{figure}

\newpage 

In sharp contrast to the compact source of hard X-rays in low-$\Mdot $
systems, there is strong evidence for an extended source of EUV/soft X-rays
in high-$\Mdot $ systems. {\it EXOSAT\/} light curves of OY~Car ($i\approx
83^\circ $) in super\-outburst (Naylor et~al.\ 1988) and {\it ROSAT\/} light
curves of the novalike variable UX~UMa ($i\approx 71^\circ $; Wood et~al.\
1995b) rule out an eclipse by the secondary of a compact source of X-rays
centered on the white dwarf. In both cases, it is argued that the boundary
layer is obscured from view by the accretion disk at all orbital phases, and
that the observed X-rays come from a corona or wind above the disk.

The situation is far more complex in U~Gem ($i\approx 70^\circ $). In
quiescence, the hard X-rays are partially eclipsed at orbital phase $\phi
\sim 0.7$ (Szkody et~al.\ 1996), and during outburst the EUV/soft X-rays are
partially eclipsed at orbital phases $\phi\sim 0.7$ and 0.1 (Mason et~al.\
1988; Long et~al.\ 1996). These dips in the \hbox{X-ray} and EUV light curves
are interpreted as due to partial eclipses of the boundary layer by vertical
structure at the edge of the accretion disk ($r\approx 4\times 10^{10}$~cm)
or at the circularization radius of the accretion stream ($r\approx 5\times
10^9$~cm); in either case, the heights above the orbital plane are
distressingly large: $h\approx 0.5\, r$, whereas the disk thickness $H\approx
c_{\rm s}/\Omega_{\rm K}\ll r$. The {\it EUVE\/} spectra of U~Gem in outburst
supply additional diagnostic information.  The phase-averaged spectrum is a
melange of Ne VI--VIII, Mg VI--VII, and Fe VII--X emission lines superposed on
a $kT\approx 12$~eV blackbody continuum (Long et~al.\ 1996). The phase-resolved
spectra demonstrate that the eclipses affect the continuum more strongly than
the lines (most strikingly, the Ne~VIII $\lambda 88.1$ line is nearly unaffected
by the eclipses; see Mauche 1997b), implying that the lines are produced in a
region of larger extent than that of the continuum, which presumably is formed
in the boundary layer.

\begin{figure}
\epsfbox[130 530 475 700]{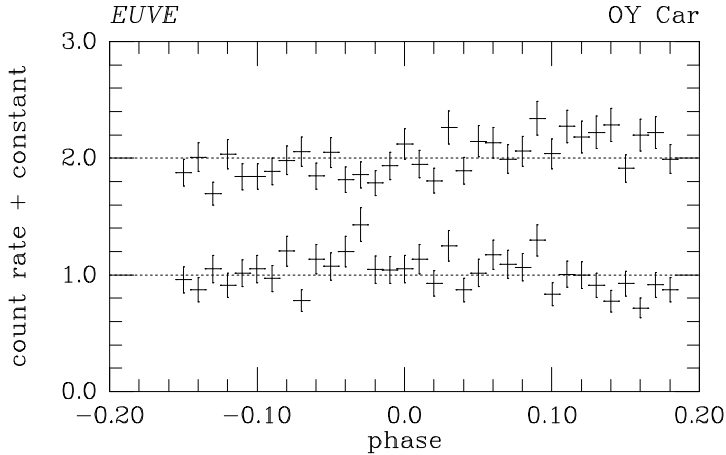}	
\caption{Typical {\it EUVE\/} DS light curves of OY~Car in super\-outburst.
The intervals shown are centered on $\rm JD-2450000=535.85$ ({\it lower\/})
and 537.49 ({\it upper\/}) and the unscaled count rates are 0.68 and 0.81
$\rm counts~s^{-1}$, respectively.}
\label{fig2}
\end{figure}

To further investigate the nature of the EUV/soft X-ray flux of OY~Car, we
recently obtained {\it EUVE\/} spectra (\S 4) and light curves of this dwarf
nova in super\-outburst. The observations were continuous for an interval of
$\approx 3$ days near the beginning of the outburst, but useful data is
obtained by {\it EUVE\/} for only $\approx 30$ min of every 94.9 min satellite
orbit. Given the 90.9 min orbital period of OY~Car, the binary phases advance
by only 4\% per satellite orbit, so data near the eclipse of the white dwarf
is obtained for only 7--8 continuous orbits every $\approx 40$ hrs.
Nevertheless, the larger number of eclipses observed (13 vs.\ 2) and the
higher count rate (0.8 vs.\ $0.02~\rm counts~s^{-1}$) lead to a significant
improvement over the previous {\it EXOSAT\/} observations. Typical {\it EUVE\/}
Deep Survey (DS) light curves of OY~Car are shown in Figure~2. As during the
{\it EXOSAT\/} observations, no eclipses are observed in any of the {\it
EUVE\/} light curves, implying that the EUV emission region is significantly
extended: the size of the emission region must be larger than that of the
secondary ($R_{\rm sec}\approx 1\times 10^{10}$~cm) and at least comparable
to the size of the binary ($a\approx 5\times 10^{10}$~cm).

\section{Dwarf Nova Oscillations}

As is well known, CVs manifest a variety of rapid periodic and aperiodic
oscillations in their optical and X-ray flux (Warner 1995). The so-called
``dwarf nova oscillations'' (DNOs; Patterson 1981) of high-$\Mdot $ CVs have
high coherence ($Q\approx 10^4$--$10^6$), periods of $\approx 10$--30~s,
amplitudes of $\approx 10$--30\% in soft X-rays and $\lax 0.5$\% in the optical,
and are sinusoidal to high accuracy. X-ray DNOs have been best studied in
SS~Cyg (C\'ordova et~al.\ 1980; 1984; Jones \& Watson 1992), U~Gem (C\'ordova
et~al.\ 1984; Mason et~al.\ 1988; Long et~al.\ 1996), and VW~Hyi (van der Woerd
et~al.\ 1987). Of these, the DNOs of SS~Cyg have been studied most extensively,
thanks to numerous {\it HEAO-1\/}, {\it EXOSAT\/}, and {\it EUVE\/} pointings of
this dwarf nova in outburst. A strong anti-correlation between the oscillation
period and the EUV luminosity (hence, by inference, $\Mdot $ onto the white
dwarf) has been discovered (Mauche 1996a), the EUV spectrum of the oscillations
has been measured (Mauche 1997a), and stringent limits have been placed on the
amount of power in higher and lower harmonics of the fundamental (Mauche
1997c). Recently, a new and surprising property of the X-ray DNOs of SS~Cyg has
been discovered.

SS~Cyg was observed in outburst in 1996 October and December by {\it EUVE\/}
and {\it ROSAT\/}, respectively. The optical and EUV light curves of the short
asymmetric 1996 October outburst are shown in the top panel of Figure~3. The
figure demonstrates the well-known lag on the rise to dwarf nova outbursts
between the optical and shorter wavelength flux (for additional information,
see Mauche \& Mattei 1997). The bottom panel of Figure~3 shows the evolution
with time of the period of the EUV oscillation. During the rise to outburst,
the period of the oscillation fell from 7.8~s to 6.6~s, jumped {\it
discontinuously} (``tunneled'') to 2.90~s, and then fell to 2.85~s over an
interval of 4.8~hr before observations with the {\it EUVE\/} DS were terminated.
When observations with the DS instrument resumed on the decline from outburst,
the period was 6.7~s and rose to 8.2~s over an interval of 2.1 days. During
the {\it ROSAT\/} observations of the plateau phase of the subsequent long
asymmetric 1996 December outburst, the period of the oscillation was observed
to rise from 2.80~s to 2.92~s over an interval of 4.2 days (van Teeseling
1997b).

\newpage 

\begin{figure}[h]
\epsfbox[130 360 475 700]{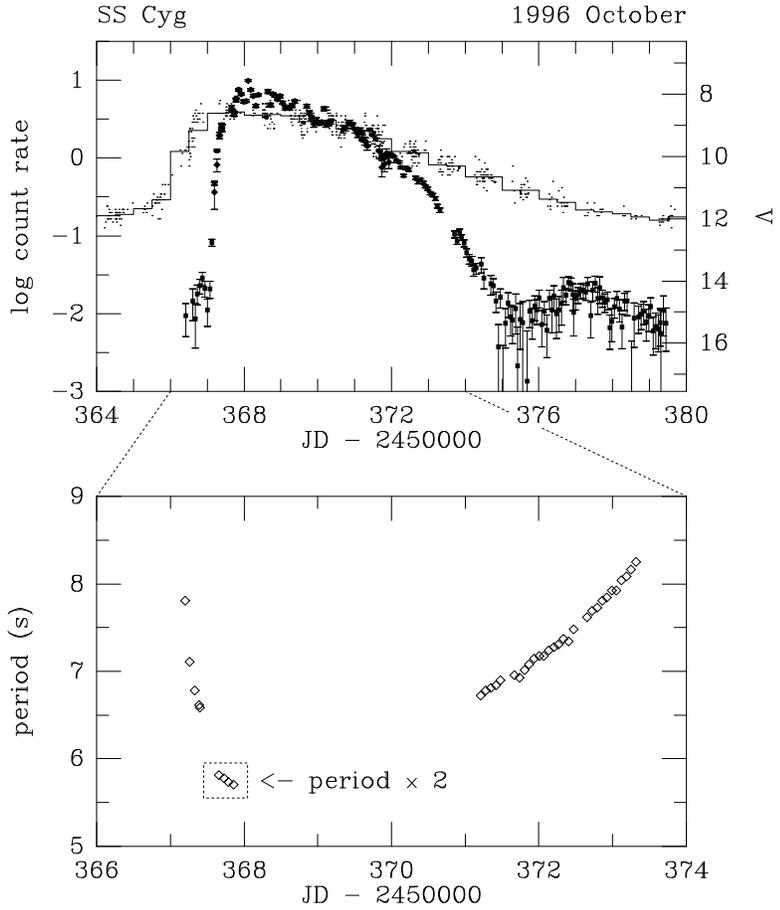}	
\caption{{\it Upper panel\/}: Optical and EUV light curves of the 1996
October outburst of SS~Cyg. {\it EUVE\/} DS and scaled SW spectrometer count
rates are shown by the filled symbols with error bars; individual AAVSO
measurements are shown by the small dots; 0.5 day mean optical light curve
is shown by the histogram. {\it Lower panel\/}: Oscillation period as a
function of time.}
\label{fig3}
\end{figure}

These observations set two precedents for DNO research: (1) periods below 7~s
and (2) period doubling. While period doubling has never been observed before
in the optical or X-ray wavebands, its ``seeds'' may have been observed
in {\it EUVE\/} observations of the long asymmetric 1994 June/July outburst of
SS~Cyg (Mauche 1997c). During the optical plateau phase of the outburst, the
amplitude of the first harmonic relative to the fundamental was 11\%, whereas
during the peak of the outburst, it was 30\%. Perhaps at even higher EUV
luminosities (hence, by inference, higher $\Mdot $ onto the white dwarf), the
power switches entirely into the first harmonic. That the $\sim 3$~s period is
the first harmonic of the $\sim 6$~s fundamental is suggested by the smooth
extrapolation of oscillation period with time shown in the lower panel of
Figure~3. However, it is also possible to consider the $\sim 3$~s period the
fundamental and the $\sim 6$~s period the first subharmonic. The Keplerian
period at the surface of the white dwarf in SS~Cyg is $P=3.76\, (\Rwd/3.85
\times 10^8~{\rm cm})^{3/2} (\Mwd/ 1.2\,\Msun)^{-1/2}$~s. For the Keplerian
period to equal the observed minimum period of 2.80~s, the mass of the white
dwarf must be $\Mwd = 1.27\,\Msun $. It would be very useful to firmly
confirm or exclude this value by independent means. Furthermore, it would be
helpful to have data on the upward {\it and\/} downward transitions of the
period doubling, to obtain simultaneous optical and/or UV data, and to
determine if this behavior is manifested by other dwarf novae. An important
question to consider is whether period doubling has never been observed in
the optical simply because the typical integration times were too long.

\section{Spectra}

Many of the spectroscopic aspects of the X-ray emission of nonmagnetic CVs
are discussed by Mauche (1997b) and consequently will not be reproduced here.
We limit the discussion here to (1) the low boundary layer luminosities of
high-$\Mdot $ systems, (2) the unique nature of the EUV/soft X-ray emission of
SS~Cyg, and (3) the EUV spectrum of OY~Car.

One of the basic predictions of theoretical models of (thin) disk accretion
onto a central object is that the ratio of the boundary layer to accretion disk
luminosity is of order unity unless the central object is rotating near breakup.
From \S 1, the ratio $L_{\rm bl}/L_{\rm disk} = [1-(v/v_{\rm break})]^2\equiv
\zeta $. From the inferred white dwarf rotation rates, $\zeta > 0.92$ for
U~Gem, $\approx 0.90$ for SS~Cyg, and $\approx 0.64$ for VW~Hyi (Mauche 1997c).
In sharp contrast, $L_{\rm bl}/L_{\rm disk}\approx 0.45$ for U~Gem (Long 
et~al.\ 1996), $\lax 0.07$ for SS~Cyg (Mauche et~al.\ 1995), and $\sim 0.04$
for VW~Hyi (Mauche et~al.\ 1991; Mauche 1996b). While we must acknowledge that
our understanding of the intrinsic spectral energy distribution of the accretion
disk {\it and\/} boundary layer is far from certain (and hence the assumed
bolometric corrections are far from certain), it becomes increasingly clear
that the boundary layer luminosities of high-$\Mdot $ CVs often fall far
short of the values predicted by simple theory. While this result is seized on
with relief by wind modelers (e.g., Drew 1997), it remains for boundary layer
modelers a serious unsolved theoretical problem.

Another apparent discrepancy between theory and observations is the effective
temperature of the boundary layer. Theory predicts $kT_{\rm eff} \approx
17$--30 eV (\S 1), but among the many high-$\Mdot $ nonmagnetic CVs detected
during the {\it ROSAT\/} All-Sky Survey (Beuermann \& Thomas 1993; Verbunt
et~al.\ 1997), only SS~Cyg in outburst displayed such a soft spectral component.
The boundary layer of U~Gem is too soft ($kT\approx 12$ eV; Long et~al.\ 1996)
and that of VW~Hyi is too dim and too soft ($kT\lax 10$ eV; Mauche et~al.\
1991; Mauche 1996b) to be detected by {\it ROSAT\/}. Recklessly assuming that
the other high-$\Mdot $ nonmagnetic CVs suffer from  similar inadequacies,
SS~Cyg in outburst stands out as unique in manifesting a relatively hard
($kT\approx 20$--30 eV; Mauche et~al.\ 1995; Ponman et~al.\ 1995) soft X-ray
spectral component.

\begin{figure}
\epsfbox[ 95 540 440 700]{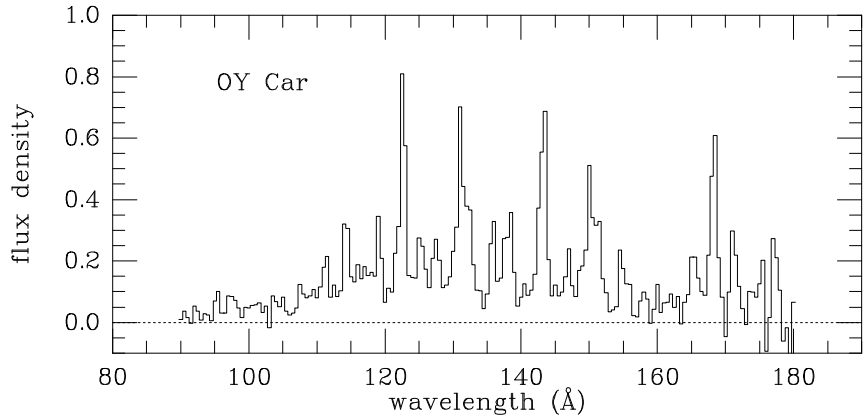}	
\caption{{\it EUVE\/} spectrum of OY~Car in super\-outburst. Units
of flux density are $10^{-12}~\rm erg~cm^{-2}~s^{-1}~\AA ^{-1}$.}
\label{fig4}
\end{figure}

In \S 2 we argued that the absence of eclipses in the {\it EXOSAT\/} and {\it
EUVE\/} light curves of OY~Car in super\-outburst evidence an extended and
hence optically thin EUV/soft X-ray emission region. That this region is indeed
optically thin is demonstrated by the EUV spectrum accumulated during the {\it
EUVE\/} observations (Fig.~4). Superposed on a very weak continuum are emission
lines of such species as O~V--VI, Ne~V--VI, Mg~IV--VI, and Fe~VI--X; species
which dominate in collisionally ionized gas at $T\approx 2$--$10\times 10^5$~K.
Of the existing {\it EUVE\/} spectra of high-$\Mdot $ CVs, this spectrum looks
most like that of U~Gem in outburst (Long et~al.\ 1996; Mauche 1997b), but it
has a significantly weaker and, with no flux shortward of $\sim 100$~\AA ,
cooler continuum. That the emission line region of OY~Car is also cooler than
that of U~Gem is indicated by the absence of the Ne~VII $\lambda 97.5$ and
Ne~VIII $\lambda 88.2$ emission lines prominent in the EUV spectrum of U~Gem.
A Mewe et~al.\ (1985) coronal model does a very poor job of reproducing the
observed spectrum, but without additional work it cannot be completely excluded
that the poor match is due simply to incomplete atomic data. A much better
match is made with a model using the Verner et~al.\ (1996) list of permitted
resonance lines. If these preliminary conclusions survive detailed analysis,
they argue that resonant scattering of the boundary layer continuum and not
thermal emission from a shocked wind is the cause of the observed EUV flux of
OY~Car (Raymond \& Mauche 1991).

\acknowledgments

We are grateful to K.~Mukai for kindly supplying the data shown in Fig.~1.
This work was performed under the auspices of the U.S.\ Department of Energy
by Lawrence Livermore National Laboratory under contract No.\ W-7405-Eng-48.



\end{document}